\documentclass[twocolumn,trackchanges]{aastex701}

\usepackage{amsmath}

\usepackage{soul}
\begin{document}

\title{Mass measurements of the double neutron star system PSR J0641+0448}

\author{Z. L. Yang}
\affiliation{National Astronomical Observatories, Chinese Academy of Sciences, Jia-20 Datun Road, ChaoYang District, Beijing 100012, China}
\affiliation{State Key Laboratory of Radio Astronomy and Technology, Beijing 100101, China }
\email{zlyang@nao.cas.cn}

\author[0000-0002-9274-3092]{J.~L. Han} %\thanks{E-mail: hjl@nao.cas.cn}
\affiliation{National Astronomical Observatories, Chinese Academy of Sciences, Jia-20 Datun Road, ChaoYang District, Beijing 100012, China}
\affiliation{School of Astronomy and Space Science, University of Chinese Academy of Sciences, Beijing 100049, China}
\affiliation{State Key Laboratory of Radio Astronomy and Technology, Beijing 100101, China }
\email[show]{E-mail: hjl@nao.cas.cn}  

\author{P.~F. Wang}
\affiliation{National Astronomical Observatories, Chinese Academy of Sciences, Jia-20 Datun Road, ChaoYang District, Beijing 100012, China}
\affiliation{State Key Laboratory of Radio Astronomy and Technology, Beijing 100101, China }
\email{E-mail: pfwang@nao.cas.cn}  

\author{C. Wang}% $^{1,2}$,
\affiliation{National Astronomical Observatories, Chinese Academy of Sciences, Jia-20 Datun Road, ChaoYang District, Beijing 100012, China}
\affiliation{School of Astronomy and Space Science, University of Chinese Academy of Sciences, Beijing 100049, China}
\affiliation{State Key Laboratory of Radio Astronomy and Technology, Beijing 100101, China }
\email{E-mail: wangchen@nao.cas.cn}

\author{N.~N. Cai}%$^{1,2}$,
\affiliation{National Astronomical Observatories, Chinese Academy of Sciences, Jia-20 Datun Road, ChaoYang District, Beijing 100012, China}
\affiliation{State Key Laboratory of Radio Astronomy and Technology, Beijing 100101, China }
\email{E-mail: nncai@nao.cas.cn}  

\author[0000-0002-1056-5895]{W.~C. Jing}
\affiliation{National Astronomical Observatories, Chinese Academy of Sciences, Jia-20 Datun Road, ChaoYang District, Beijing 100012, China}
\affiliation{School of Astronomy and Space Science, University of Chinese Academy of Sciences, Beijing 100049, China}
\email{E-mail: wcjing@nao.cas.cn}  

\author{W.~Q. Su}%$^{1,2}$,
\affiliation{National Astronomical Observatories, Chinese Academy of Sciences, Jia-20 Datun Road, ChaoYang District, Beijing 100012, China}
\affiliation{State Key Laboratory of Radio Astronomy and Technology, Beijing 100101, China }
\email{E-mail: wqsu@nao.cas.cn}  

\author{T. Wang}
\affiliation{National Astronomical Observatories, Chinese Academy of Sciences, Jia-20 Datun Road, ChaoYang District, Beijing 100012, China}
\affiliation{State Key Laboratory of Radio Astronomy and Technology, Beijing 100101, China }
\email{E-mail: twang18@nao.cas.cn}  

\author{J. Xu}
\affiliation{National Astronomical Observatories, Chinese Academy of Sciences, Jia-20 Datun Road, ChaoYang District, Beijing 100012, China}
\affiliation{State Key Laboratory of Radio Astronomy and Technology, Beijing 100101, China }
\email{E-mail: xujun@nao.cas.cn}  

\author{Yi Yan}
\affiliation{National Astronomical Observatories, Chinese Academy of Sciences, Jia-20 Datun Road, ChaoYang District, Beijing 100012, China}
\affiliation{State Key Laboratory of Radio Astronomy and Technology, Beijing 100101, China }
\email{E-mail: yanyi@nao.cas.cn}  

\author{D.~J. Zhou} %$^{1,2}$,
\affiliation{National Astronomical Observatories, Chinese Academy of Sciences, Jia-20 Datun Road, ChaoYang District, Beijing 100012, China}
\affiliation{State Key Laboratory of Radio Astronomy and Technology, Beijing 100101, China }
\email{E-mail: djzhou@nao.cas.cn}

\begin{abstract}

Pulsar timing of double neutron star (DNS) systems is one of the best  methodologies to study the neutron star masses distribution. Here we report the discovery of a double neutron star system PSR J0641+0448 in the Five-hundred-meter Aperture Spherical radio Telescope (FAST) Galactic Plane Pulsar Snapshot (GPPS) survey. This pulsar has a 25.7 ms spin period and moves in a 3.73-days eccentric orbit with an eccentricity of 0.145. Using FAST observations, we obtained its phase-connected timing solution with periastron advance and Shapiro delay detected. Using $\chi^2$ analysis based on DDGR model, we constrain the pulsar mass to $1.319^{+0.021}_{-0.035}~M_\odot$, and the companion mass to $1.269^{+0.022}_{-0.016}~M_\odot$ with a 68.3\% confidence level.  
The low companion mass and mild orbital eccentricity is consistent with the correlation between neutron masses and orbital eccentricities. 

\end{abstract}



\section{Introduction} 

The birth mass distribution of neutron stars represents a fundamental problem in binary pulsar population synthesis \citep{Tauris+2023pbse.book.....T} and carries vital information about the supernova (SN) explosion mechanism \citep{Pejcha+2012MNRAS.424.1570P}. In binary systems, neutron star masses can increase through mass transfer processes such as wind accretion, Roche-lobe overflow, and common envelope evolution \citep{Tauris+2023pbse.book.....T}. Most binary pulsars are observed with low-mass companions, including He white dwarfs and low-mass main-sequence stars \citep{Manchester+2005AJ....129.1993M}. These systems descend from low-mass X-ray binaries, in which mass transfer proceeds over extended timescales, often leading to substantial mass growth of the pulsar \citep{Tauris+2023pbse.book.....T}. By contrast, DNS systems experience only short-lived mass transfer episodes, yielding minimal increase in neutron star mass \citep{Tauris+2017ApJ...846..170T}. As a result, the mass of the first-born neutron star remains close to its birth value, while the second-born neutron star retains its original birth mass entirely. Given that, the mass distribution of Galactic DNS systems has been extensively investigated to constrain the birth mass distribution of neutron stars \citep{Ozel+2012ApJ...757...55O,Ozel+2016ARA&A..54..401O,You+2025NatAs...9..552Y}. 

The potential to determine the masses of the pulsar and its companion through pulsar timing was recognized immediately after the discovery of the first DNS system \citep{Hulse+1975ApJ...195L..51H}. These systems exhibit eccentric orbits and contain massive companions, which significantly facilitate the detection of relativistic effects. Such relativistic effects are described using post-Keplerian (PK) parameters \citep{Damour+1985AIHPA..43..107D,Damour+1986AIHS...44..263D}. When two or more PK parameters are measured, the component masses of the system can be determined within the framework of general relativity \citep{Damour+1992PhRvD..45.1840D}. Three or more PK parameters enable rigorous tests of gravitational theories, as demonstrated with exceptional precision in the double pulsar system PSR J0737--3039A/B \citep{Kramer+2021PhRvX..11d1050K}. So far, approximately 60 neutron stars have accurately measured masses\footnote{https://www3.mpifr-bonn.mpg.de/staff/pfreire/NS\_masses.html\label{NSmass}}, many of which are also members of DNS systems. Despite that, yet a larger catalog of neutron star mass measurements remains essential.

\begin{table*}[tb]
\centering
\caption{{\bf Log of FAST observations of PSR~J0641+0448.}  Listed are the Gregorian observation date, cover name jointed with FAST beam name, FAST project number,  project principal investigator (PI), and observation length in minutes for each observation session.  }
%\setlength{\tabcolsep}{2.5pt}
%\fontsize{6}{7.5}\selectfont
\footnotesize
    \begin{tabular}{cllcc}
    \hline
     Obs. Date & Cover \& Beam & Project ID & PI & Length  \\
     (yyyymmdd)     &     &    &      & (min)        \\
     \hline
     20240516 & G207.62+0.00M05P1 & ZD2023\_2 & J.L. Han & 5 \\
     20240608 & J064133+044835M01 & ZD2023\_2 & J.L. Han & 15  \\
     20240629 & J064133+044835M01 & ZD2023\_2 & J.L. Han & 15  \\
     20240811 & J0641+0448gM01 & ZD2024\_2 & J.L. Han & 5  \\
     20250101 & J0641+0448gM01 & ZD2024\_2 & J.L. Han & 10  \\
     20250205 & J0641+0448gM01 & ZD2024\_2 & J.L. Han & 5  \\
     20250308 & J0641+0448gM01 & ZD2024\_2 & J.L. Han & 5  \\
     20250612 & J0641+0448gM01 & ZD2024\_2 & J.L. Han & 5  \\
     20250812 & J0641+0448gM01 & PT2025\_0179 & C. Wang & 5  \\
     20250815 & J0641+0448gM01 & ZD2025\_2 & J.L. Han & 5  \\
     20250822 & J0641+0448M01 & PT2025\_0150 & P.F. Wang & 11  \\
     20250915 & J0641+0448M01 & PT2025\_0150 & P.F. Wang & 10  \\
     20250930 & J0641+0448M01 & PT2025\_0150 & P.F. Wang & 15  \\
     20251014 & J0641+0448M01 & PT2025\_0161 & P.F. Wang & 10  \\
     20251104 & J0641+0448M01 & PT2025\_0161 & P.F. Wang & 10  \\
     20251119 & J0641+0448M01 & PT2025\_0161 & P.F. Wang & 10\\
     20251205 & J0641+0448M01 & PT2025\_0161 & P.F. Wang & 4\\
     20251211 & J0641+0448gM01 & ZD2025\_2 & J.L. Han & 10\\
     20251215 & J0641+0448gM01 & ZD2025\_2 & J.L. Han & 10\\
     20251217 & J0641+0448gM01 & PT2025\_0179 & C. Wang & 5\\
     20251219 & J0641+0448gM01 & ZD2025\_2 & J.L. Han & 10\\
     20260101 & J0641+0448gM01 & ZD2025\_2 & J.L. Han & 10\\
     20260105 & J0641+0448gM01 & ZD2025\_2 & J.L. Han & 10\\
     20260205 & J0641+0448gM01 & ZD2025\_2 & J.L. Han & 10\\
     20260216 & J0641+0448gM01 & ZD2025\_2 & J.L. Han & 10\\
    \hline
     \end{tabular}
     \label{obsinfo}
\end{table*}

Here we report the discovery of a new double neutron star (DNS) system,  J0641+0448, identified by the Five-hundred-meter Aperture Spherical radio Telescope \citep[FAST;][]{Nan+2006ScChG..49..129N,Nan+2011IJMPD..20..989N} through its Galactic Plane Pulsar Snapshot (GPPS) survey \citep{Han+2021RAA....21..107H}. The GPPS survey aims to discover pulsars within $\pm 10^\circ$ of the Galactic latitude from the Galactic plane in the sky visible to FAST and has so far detected over 800 new pulsars \citep{Han+2021RAA....21..107H,Han+2025RAA}, including three DNS systems: PSRs J0528+3529, J1844--0128, and J1901+0658 \citep{Su+2023MNRAS,wang+2025RAA}. PSR~J0641+0448 is a mildly recycled pulsar with a spin period $P = 25.7~\mathrm{ms}$, moving in a $3.73$-day eccentric orbit ($e = 0.145$). The mass function implies a minimum companion mass of $1.28~M_\odot$, indicating a massive companion and supporting the classification as a DNS system. Using more than  600 days of FAST observations, we have obtained a phase-connected timing solution and constrained the component masses of the system.

The structure of this paper is as follows: Section~\ref{sec2} describes the discovery and follow-up observations of PSR~J0641+0448; Section~\ref{sec3} presents the timing results and mass measurements; and Section~\ref{sec4} provides discussion and conclusion of our findings.

\begin{figure}
    \centering
    \includegraphics[width=0.85\columnwidth]{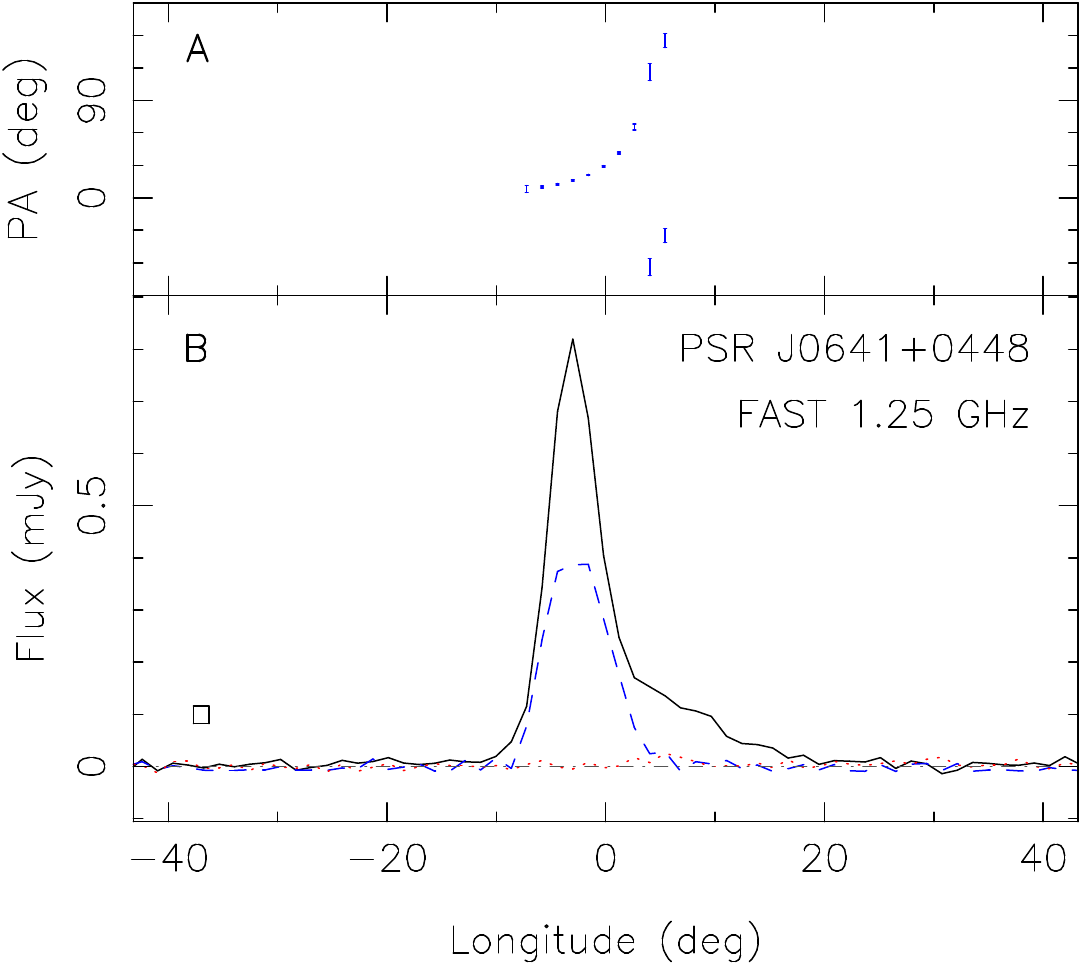}
    \caption{Polarization profiles of PSR J1856--0039, integrated over 13 hours of observations. Panel (A): Position angles (PAs) of the linear polarization at infinite frequency. Error bars represent $\pm1\sigma$ uncertainties. Panel (B): Total intensity $I$ (black solid line), linear polarization $L$ (blue dashed line), and circular polarization $V$ (red dotted line, with positive values corresponding to left-hand circular polarization). The small square in the bottom left corner indicates the $\pm2\sigma_{\rm I,off}$ noise level in total intensity and a width of one phase bin.}
    \label{pol}
\end{figure}

\begin{figure}
    \centering
    \includegraphics[width=0.85\columnwidth]{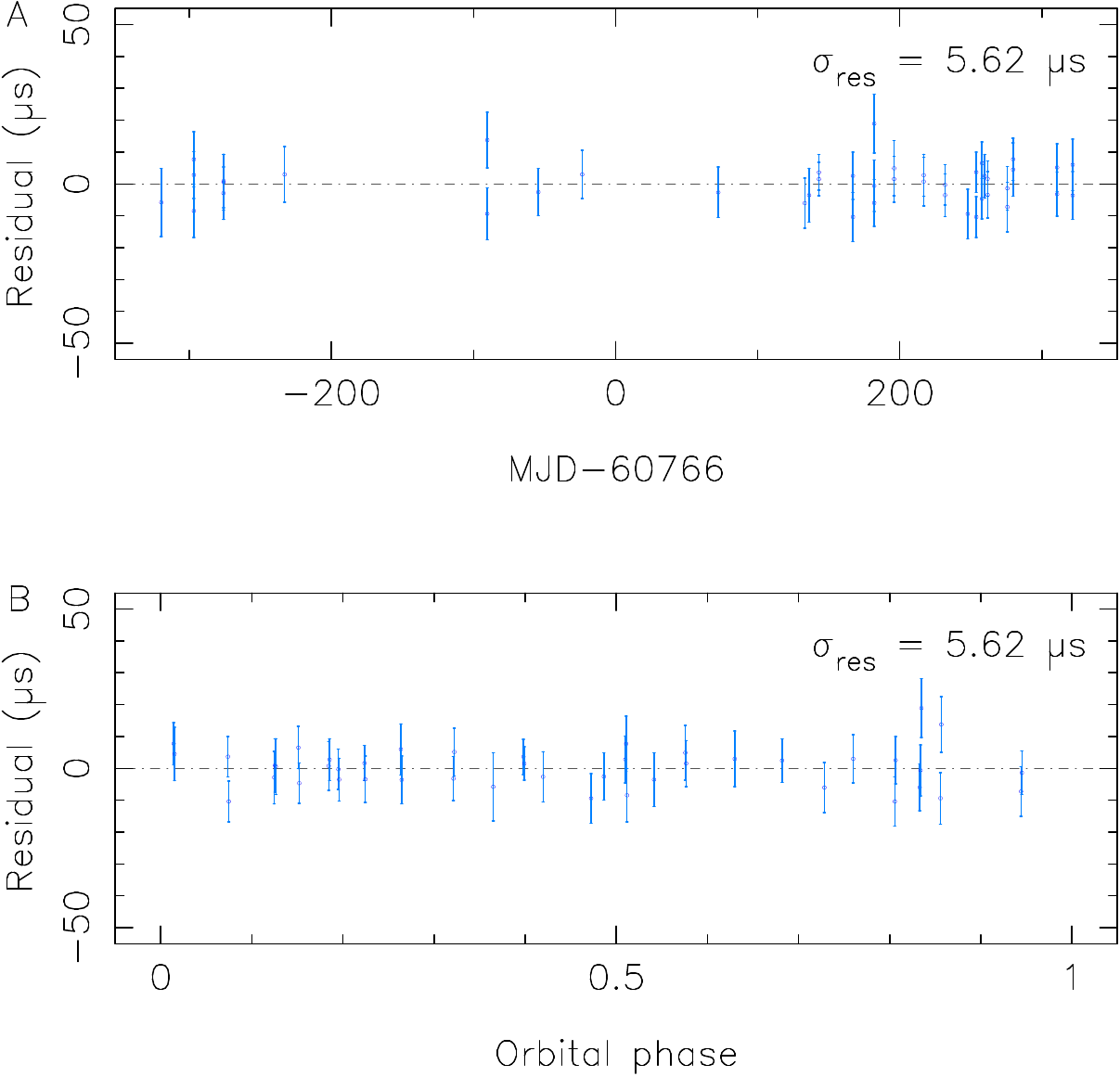}\\[3mm]
    \includegraphics[width=0.85\columnwidth]{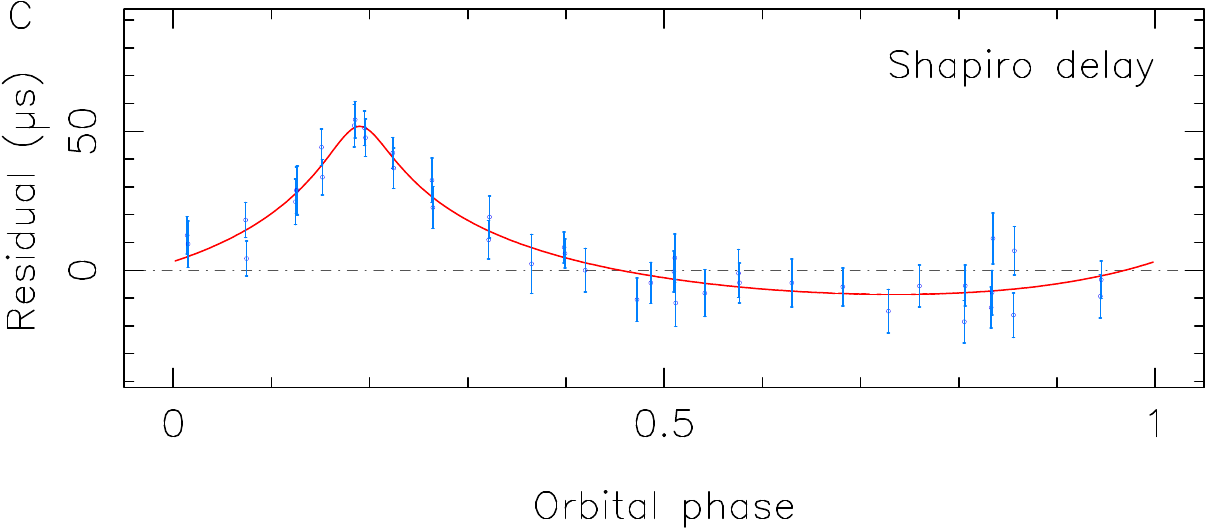}
    \caption{Timing residuals for PSR~J0641+0448. The post-fit timing residuals are plotted versus epoch and orbital phase in subpanels (A) and (B), respectively. No systematic trends are apparent. The weighted root-mean-square (RMS) of the residuals, $\sigma_{\rm res}$, is 5.62 \textmu s. Subpanel (C) shows the Shapiro delay predicted by the best-fit values of DDGR binary model ($M_{\rm c}=1.269~M_\odot,i=79^\circ.6$).}
    \label{timeres}
\end{figure}

\section{FAST Observations and Data Reduction}
\label{sec2}

\subsection{FAST observations of PSR~J0641+0448}

PSR J0641+0448 \cite[GPPS No. 0670, originally designated PSR~J0641+0448g in][]{Han+2025RAA} was first detected in an 88‑minute SnapShotZCal observation of the FAST GPPS survey on 2024 May 16. The detection was confirmed by a 15‑minute tracking observation on 2024 June 8. To date, a total of 22 observational sessions have been carried out with FAST to study this source which are listed in Table~\ref{obsinfo}. All observations used the $L$-band 19‑beam receiver, covering a frequency range from 1.0 to 1.5 GHz with a sampling time of 49.152 \textmu s. The initial observation was recorded with 1024 frequency channels and 2 polarization channels ($XX$ and $YY$), while all subsequent observations utilized 2048 frequency channels and 4 polarization channels ($XX$, $YY$, $\text{Re}[X^{}Y]$, and $\text{Im}[X^{}Y]$). As part of the standard calibration procedure, a 2.01326~s period, 1.1~K modulated signal is injected into the receiver at either the beginning or the end of each FAST observation session.

\subsection{Pulsar timing and polarization profiles}

We derived the barycentric spin period of PSR~J0641+0448 at different barycentric epochs from each observation using \textsc{Presto} \citep{Ransom+2001PhDT.......123R}. The variation in the barycentric spin period indicates that the pulsar is in a binary system. Following the procedure described in \citet{wang+2025RAA}, which is based on the method introduced by \citet{Bhattacharyya+2008MNRAS.387..273B}, we estimated the initial orbital parameters using the first eight observations.

Using these orbital parameters, we folded the raw \textit{psrfits} data into pulse profiles with \textsc{DSPSR} \citep{Straten+2011PASA...28....1V}, adopting a sub-integration length of 30~s. The resulting profiles were then calibrated and cleaned using the \textsc{Psrchive} software package \citep{Hotan+2004PASA...21..302H}. Under the ideal feed assumption, we folded the calibration signals and applied the corresponding calibration to the pulsar profiles using the \texttt{pac} command. Radio frequency interference was excised manually with the \texttt{psrzap} tool.

Subsequently, we reduced the data volume by averaging the full frequency channels into 4 sub-bands and the sub-integrations into 300~s bins using the \texttt{pam} command. TOAs were extracted from these integrated profiles with the \texttt{pat} command by cross-correlating them with a frequency-averaged 1D template generated by \texttt{paas}. The dispersion measure (DM) value was refined by comparing TOAs from different sub-bands. Then we summed all frequency channels together and then regenerate the TOAs. The final number of TOAs is  44.
We conducted pulsar timing using the \textsc{Tempo2} software package \citep{Hobbs+2006MNRAS.369..655H}. The phase-connected timing solution for PSR~J0641+0448 was obtained following the methodology outlined by \citet{Freire+2018MNRAS.476.4794F}. The resulting timing model was then used to reprocess the observational data, yielding improved TOA measurements. The final timing solution was obtained based on these refined TOAs.

To derive the polarization profile, we first analyzed the 15-minute tracking observation from 2024 June 29 using the \texttt{rmfit} command, obtaining a Faraday rotation measure of $RM_{\rm obs} = 140.5 \pm 1.6$~rad\,m$^{-2}$. We then applied Faraday rotation correction via the \texttt{pam} command. Using the vertical total electron content maps of the ionosphere, CODE\footnote{ftp.aiub.unibe.ch/CODE/} and the International Geomagnetic Reference Field (IGRF-13)\footnote{https://www.ngdc.noaa.gov/IAGA/vmod/igrf.html} with an updated code IONFR\footnote{https://sourceforge.net/projects/ionfarrot/} \citep{Sotomayor+2013A&A...552A..58S}, we determined the ionospheric contribution to the rotation measure to be $3.8 \pm 0.3$~rad\,m$^{-2}$. The intrinsic rotation measure of the pulsar is therefore $136.7 \pm 1.6$~rad\,m$^{-2}$. Following the same procedure, we applied Faraday rotation correction to each tracking observation and subsequently combined them using the \texttt{psradd} command. The resulting integrated polarization profile, along with the position angle of linear polarization at the infinite frequency, is shown in Fig.~\ref{pol}.

\begin{table*}
\caption{Timing solution and derived parameters for PSR J0641+0448. All listed uncertainties represent the 68.3\% confidence level (C.L.).}
\setlength{\tabcolsep}{20pt}
\centering
\begin{tabular}{lc}
\hline\hline
\multicolumn{2}{c}{General Information} \\
\hline
Pulsar name & J0641+0448 \\ 
Data span (MJD) & 60446 to 61088 \\ 
Number of TOAs & 44 \\
$\chi^2/N_{\rm free}$  &  0.8407\\
Time scale  & TCB \\
Solar system ephemeris model  & DE440 \\
Binary model & DD \\
Reference epoch (MJD) & 60600\\ 
\hline
\multicolumn{2}{c}{Measured Parameters} \\ 
\hline
Right ascension (J2000) &  $06^{\rm h}41^{\rm m}34^{\rm s}.42114\pm0^{\rm s}.00008$ \\ 
Declination (J2000)  &   $04^\circ48'09''.870\pm0''.004$\\ 
Spin frequency, $\nu$ (s$^{-1}$) & 38.94359974181$\pm$0.00000000003 \\ 
Derivative of spin frequency, $\dot{\nu}$ (s$^{-2}$) & $(-9.95\pm0.05)\times10^{-17}$ \\ 
Dispersion measure, $DM$ (pc\,cm$^{-3}$) & 155.185$\pm$0.002\\
Rotation measure, $RM$ (rad\,m$^{-2}$) & 136.7$\pm$1.6 \\
Orbital period, $P_{\rm orb}$ (d) & 3.73073229$\pm$0.00000004 \\ 
Epoch of periastron, $T_0$ (MJD) &  60601.646366$\pm$0.000004 \\ 
Projected semi-major axis of orbit, $x$ (lt-s) & 15.552366$\pm$0.000011 \\ 
Longitude of periastron, $\omega$ (deg) &  5.1038$\pm$0.0004 \\ 
Orbital eccentricity, $e$ & 0.1454502$\pm$0.0000003 \\ 
Periastron advance, $\dot{\omega}$ (deg\,yr$^{-1}$) &0.0427$\pm$0.0003\\ 
Sine of orbital inclination, $\sin i$ & 0.98$\pm$0.02 \\
Companion mass, $M_{\rm c}$ ($M_\odot$) & 1.3$\pm$0.8 \\
\hline
\multicolumn{2}{c}{Derived Parameters} \\
\hline
Spin period, $P$ (s) & 0.02567816038142$\pm0.00000000000002$ \\
Spin period derivative, $\dot{P}$ (s s$^{-1}$)  & $(6.56\pm0.04)\times10^{-20}$\\
Characteristic age, $\tau_{\rm c}$ (Gyr)  & 6.2  \\
Surface magnetic field strength, $B_{\rm s}$ (G)  & $1.3\times10^{9}$ \\
Mass function, $f$ ($M_\odot$)  & 0.2901910$\pm$0.0000007 \\
DM derived distance, $d_{\rm NE2001}$ \& $d_{\rm YMW16}$ (kpc) & 5.3 \& 2.6 \\
Pulsar mass, ${M_{\rm p}}$ ($M_\odot$)$^\dagger$ & $1.319^{+0.021}_{-0.035}$ \\
Companion mass, ${M_{\rm c}}$ ($M_\odot$)$^\dagger$ & $1.269^{+0.022}_{-0.016}$ \\
Total mass, ${M_{\rm tot}}$ ($M_\odot$)$^\dagger$ & $2.588_{-0.031}^{+0.029}$ \\
Orbital inclination, $i$ (deg)$^{\dagger}$  & $79.6_{-4.1}^{+2.7}$ \\
\hline
\end{tabular}\\
$^\dagger$ Obtained via $\chi^2$ analysis based on DDGR model.
\label{timingsolution}
\end{table*}

\begin{figure*}
    \centering
    \includegraphics[width=0.85\columnwidth]{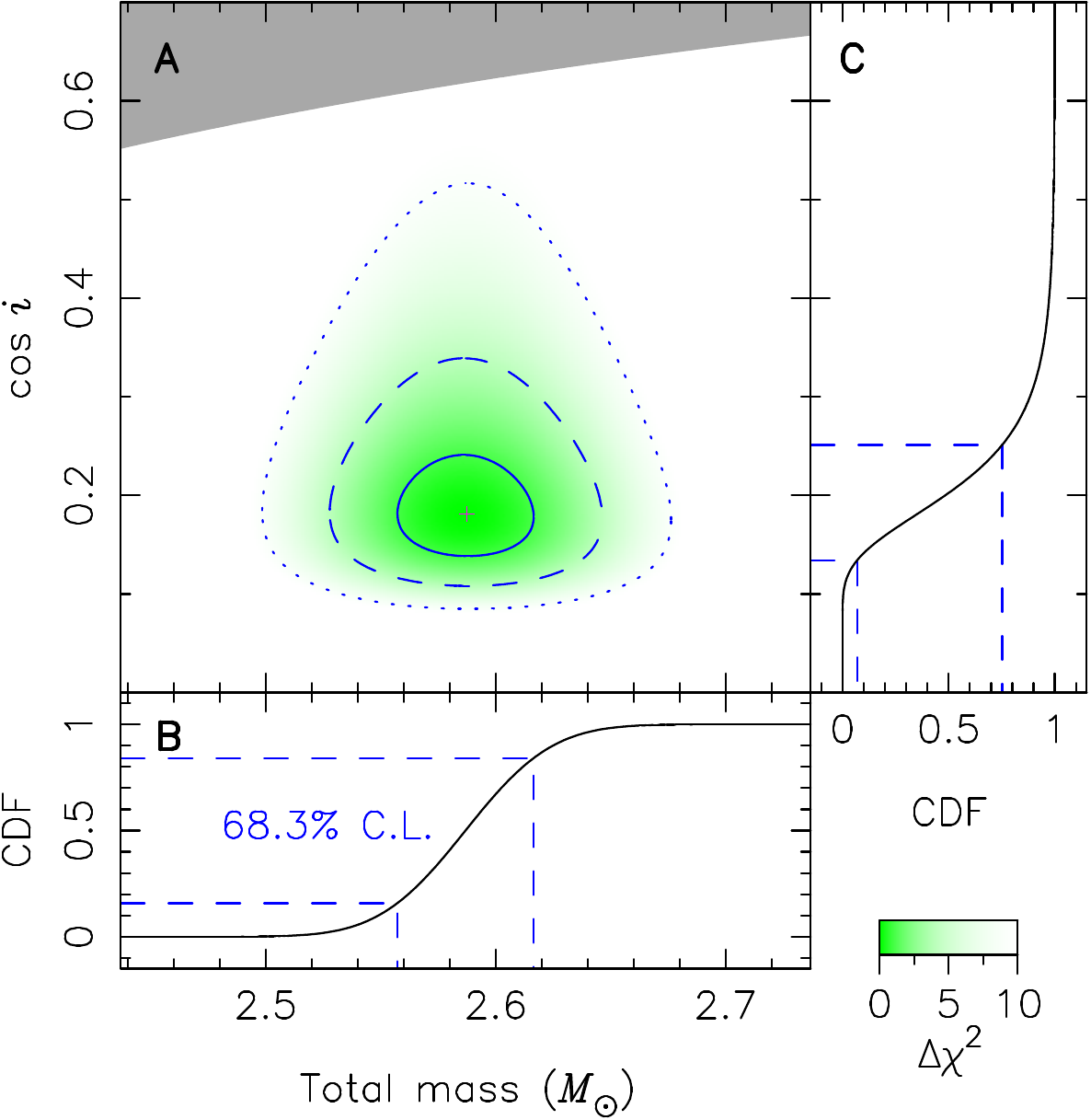} \hspace{10pt} %\\[3mm]
    \includegraphics[width=0.85\columnwidth]{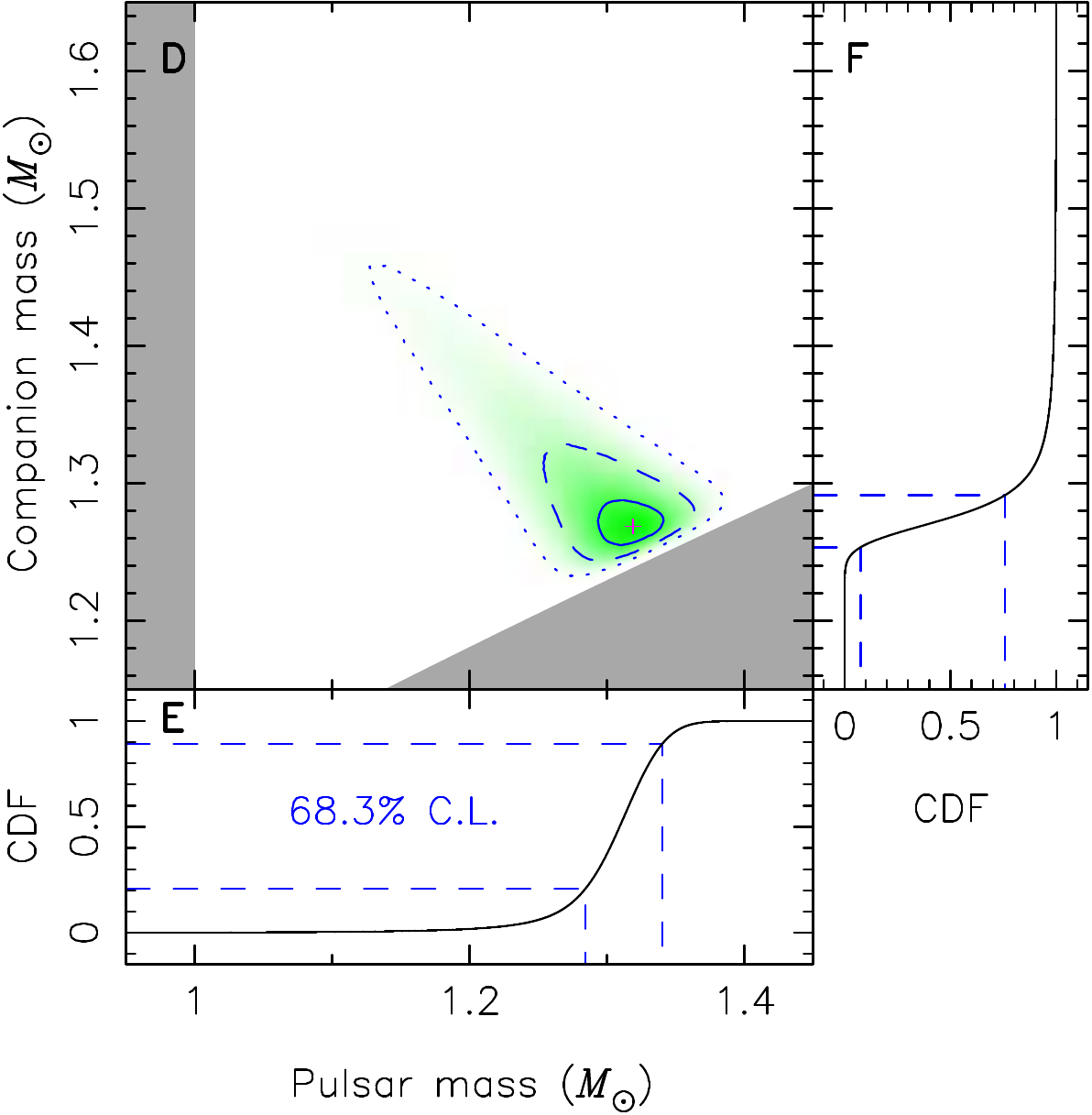}
    \caption{Constraints on the component masses and orbital inclination of PSR~J0641+0448 derived from $\chi^2$ analysis. The two‐dimensional probability distributions in the $M_{\rm{tot}}$–$\cos i$ and $M_{\rm{p}}$–$M_{\rm{c}}$ planes are shown in subpanels (A) and (D), respectively. The contours correspond to $\Delta\chi^2$ levels of 1, 4, and 9. The best‐fit values, $M_{\rm{tot}} = 2.588~M_\odot$, $i = 79^\circ.6$, $M_{\rm{p}} = 1.319~M_\odot$, and $M_{\rm{c}} = 1.269~M_\odot$ are marked by red crosses. The  highest density intervals containing 68.3\% probability (enclosed by the blue dashed lines), derived from the one‐dimensional cumulative distribution functions (indicated by the black lines) in panels (B), (C), (E), and (F), are  correspond to: $M_{\rm{tot}} = 2.557~\text{to}~2.617~M_\odot$, $i = 75.5~\text{to}~82.3$ deg, $M_{\rm{p}} = 1.284~\text{to}~1.340~M_\odot$, and $M_{\rm{c}} = 1.253 ~\text{to}~1.291~M_\odot$.}
    \label{Massmeasuremnet}
\end{figure*}

\section{Timing results and mass measurement}
\label{sec3}

\subsection{Timing result}

The timing solution for PSR~J0641+0448 is presented in Table~\ref{timingsolution}, with the corresponding timing residuals displayed in Fig.~\ref{timeres}. The absence of significant systematic trends in the residuals as a function of epoch or orbital phase indicates that the timing model provides a good fit to the data. The timing analyses are based on the Barycentric Coordinate Time (TCB) scale. We used the Solar System ephemeris DE440 \citep{Park+2021AJ....161..105P} to correct the motion of FAST to the Solar System barycenter. The orbital motion of the pulsar was modeled using the DD binary model \citep{Damour+1985AIHPA..43..107D,Damour+1986AIHS...44..263D}, which is widely adopted for DNS systems. 

The pulsar has a characteristic age of 6.2~Gyr and a surface magnetic field strength of $1.3\times10^9$~G, values that are typical of a mildly recycled pulsar. Its Galactic longitude is  $207^\circ.53441$ and Galactic latitude is $0^\circ.00106$. Using the NE2001 Galactic electron density model \citep{ne2001}, the distance derived from the dispersion measure (DM) is 5.3~kpc, whereas the YMW16 model \citep{ymw16} yields 2.6~kpc. A future measurement of the orbital period derivative, which includes kinematic contributions from proper motion and Galactic acceleration, will help to constrain the system's distance and thereby improve the Galactic electron density model in this direction.

The timing solution yields precise Keplerian orbital parameters. From these, we derive the mass function, $f(M_{\rm p},M_{\rm c},\sin i)$, defined as:
\begin{equation}
    f(M_{\rm p},M_{\rm c},i)=\frac{(M_{\rm c} \sin i)^3}{(M_{\rm p}+M_{\rm c})^2}=\frac{4\pi^2}{G}\frac{x^3}{P_{\rm b}^2},
    \label{fn}
\end{equation}
where $M_{\rm p}$ and $M_{\rm c}$ are the masses of the pulsar and its companion, respectively, $i$ is the orbital inclination angle, $G$ is the gravitational constant, $x$ is the projected semi-major axis of the pulsar's orbit, and $P_{\rm b}$ is the binary orbital period. Using the parameters from Table~\ref{timingsolution}, we obtain a mass function of $0.290\,M_\odot$. This implies a minimum companion mass of $1.28\,M_\odot$ for an assumed pulsar mass of $M_{\rm p} = 1.4\,M_\odot$.

The periastron advance of this binary pulsar has been precisely measured, and in general relativity it's related to the total mass of the system $M_{\rm tot}$ as:
\begin{equation}
    \dot{\omega}=3(\frac{P_{\rm orb}}{2\pi})^{-5/3}G^{2/3}M_{\rm tot}^{2/3}c^{-2}(1-e^2)^{-1}.
    \label{omdot}
\end{equation}
Using the $\dot{\omega}$ value and uncertainty in Table~\ref{timingsolution}, we found that the total mass of the system is $2.59\pm0.03~M_\odot$.

For eccentric orbits, the Shapiro delay is expressed as \citep{Damour+1986AIHS...44..263D}:
\begin{equation}
    \begin{split}
        \Delta_{\rm S} = -\frac{2G M_{\rm c}}{c^3} \ln \{1& - e \cos u - \sin i [ \sin \omega (\cos u - e) \\
    & + \sqrt{1 - e^2} \cos \omega \sin u ] \},
    \end{split}
\end{equation}
where $e$ is the orbital eccentricity and $u$ is the eccentric anomaly. In PSR~J0641+0448 the Shapiro delay was marginally detected (see Table~\ref{timingsolution}). 

\subsection{Mass measurement}

Due to a mild orbital eccentric of 0.145, we can assume that the spin axis of PSR~J0641+0448 is aligned with the direction of orbital angular momentum \citep{Tauris+2017ApJ...846..170T}. Then combined with the Keplerian orbital parameters in Table~\ref{timingsolution}, under general relativity all PK parameters in DD binary model can be calculated using $M_{\rm p}$ and $M_{\rm c}$. We  can combine all these relativistic effects together for the mass measurements, the realization of which is DDGR binary model \citep{Taylor+1987grg..conf..209T,Taylor+1989ApJ...345..434T}, where the fitting parameters are $M_{\rm c}$ and $M_{\rm tot}$ instead of PK parameters. Applying this binary model to PSR~J0641+0448,  \textsc{Tempo2} \citep{Hobbs+2006MNRAS.369..655H} reported $M_{\rm c}=1.269\pm0.015~M_\odot$ and $M_{\rm tot}=2.588\pm0.030~M_\odot$. These uncertainties are produced based on the approximation of linear propagation of uncertainties \citep{Hobbs+2006MNRAS.369..655H}. However, the uncertain of companion mass is significantly asymmetric, so this approximation is not enough accurate. 

Following \citet{Splaver+2002ApJ...581..509S}, we employed a $\chi^2$ analysis to better constrain the masses of PSR~J0641+0448 and its companion. A uniform prior probability distribution was adopted in the $M_{\rm tot}$–$\cos i$ plane. Since the pulsar mass $M_{\rm p}$ must exceed 1~$M_\odot$, the corresponding gray region in Fig.~\ref{Massmeasuremnet}A is excluded. For each pair of trial values $(M_{\rm tot},\cos i)$, we calculated $M_{\rm c}$ using Eq.~\ref{fn} and then fitted the TOAs with  DDGR timing model \citep{Taylor+1987grg..conf..209T,Taylor+1989ApJ...345..434T} in which all other parameters were free to vary, and recorded the resulting $\Delta\chi^2 = \chi^2 - \chi^2_{\rm min}$. Here we have rescaled the $\chi^2$ distribution so the minimum $\chi^2$ is equal to the degree of freedom. The posterior probability density $p(M_{\rm tot},\cos i)$ is then given by:
\begin{equation}
p(M_{\rm tot},\cos i)=
\begin{cases}
C e^{-\Delta\chi^2/2} & \text{if $M_{\rm p} > 1~M_\odot$}, \\
0 & \text{otherwise},
\end{cases}
\end{equation}
where $C$ is a normalization constant ensuring a total probability of unity. Using the mass function, we converted the $\Delta\chi^2$ distribution $\Delta\chi^2(M_{\rm tot},\cos i)$ into $\Delta\chi^2(M_{\rm p},M_{\rm c})$ which is shown in Fig.~\ref{Massmeasuremnet}D. The gray region has been excluded as it corresponds to an unphysical $\sin i>1$. The minimum $\chi^2$ occurs at $M_{\rm tot}=2.588~M_\odot$, $i=79^\circ.6$, $M_{\rm p}=1.319~M_\odot$, and $M_{\rm c}=1.269~M_\odot$, marked by the two red crosses in Fig.~\ref{Massmeasuremnet}A and D. Integrating $p(M_{\rm tot},\cos i)$, we obtained the following parameter ranges with a 68.3\% confidence level (C.L.):  $M_{\rm{tot}} = 2.557~\text{to}~2.617~M_\odot$, $i = 75.5~\text{to}~82.3$ deg, $M_{\rm{p}} = 1.284~\text{to}~1.340~M_\odot$, and $M_{\rm{c}} = 1.253 ~\text{to}~1.291~M_\odot$.

\section{Discussion and conclusion}
\label{sec4}

We have reported the discovery, timing solution, and mass measurements of PSR~J0641+0448, a mildly recycled pulsar with a spin period $P = 25.7$~ms, in a 3.73-day eccentric orbit ($e=0.145$) around another neutron star, identified through the FAST GPPS survey. Based on 16 FAST observing sessions, we obtained a phase-connected timing solution and detected both periastron advance and Shapiro delay. Following  \citet{Splaver+2002ApJ...581..509S}, we performed $\chi^2$ analysis based on DDGR binary model, which constrained the pulsar mass to $1.319^{+0.021}_{-0.035}~M_\odot$ and the companion mass to $1.269_{-0.016}^{+0.022}~M_\odot$.

In future, continued follow-up observations PSR~J0641+0448 can improve the precision of the mass measurement. With a longer time span of about 10 years, the Einstein delay parameters $\gamma$ and the orbital period derivative $\dot{P}_{\rm orb}$ can be measured, which can be used to test general relativity and constrain the pulsar distance.

\begin{figure}
    \centering
    \includegraphics[width=0.9\columnwidth]{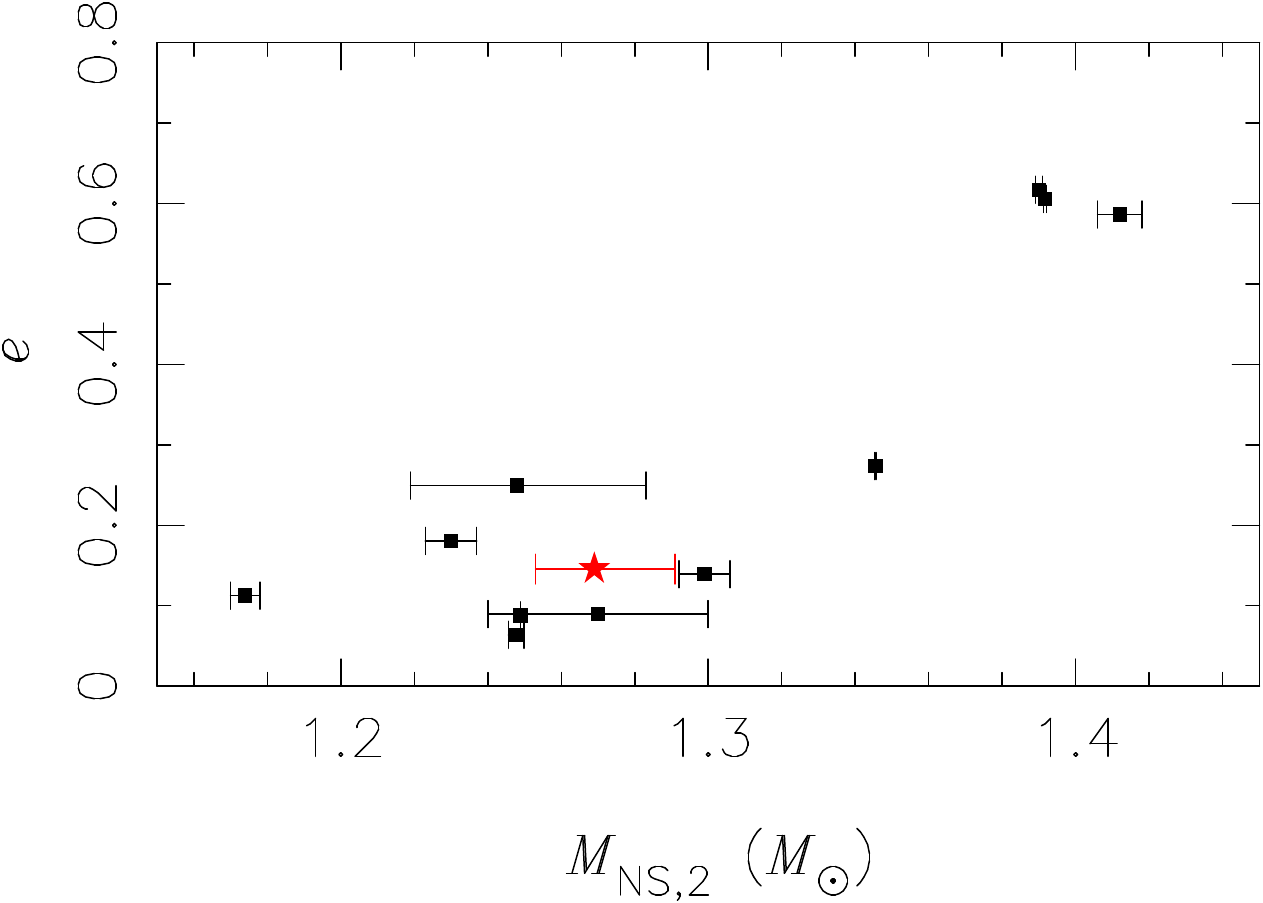}
    \caption{The masses of the second-born neutron star, $M_{\rm NS,2}$ and orbital eccentricities, $e$ of known DNS systems in the Galactic field. PSR~J0641+0448 is marked by the red star. The error bars show the $1\sigma$ uncertainties of their masses.}
    \label{MNS-ECC}
\end{figure}

It has been proposed that core-collapse events involving stellar cores with low compactness, such as electron-capture supernovae, low-mass iron core-collapse supernovae, and ultra-stripped supernovae with small metal cores, impart systematically smaller kicks than those from more massive iron-core progenitors \citep{Janka+2017ApJ...837...84J, Tauris+2017ApJ...846..170T}. Such smaller kicks are expected to yield lower orbital eccentricities, while the neutron stars produced in these events are predicted to have lower masses \citep{Ugliano+2012ApJ...757...69U, Sukhbold+2016ApJ...821...38S}. This suggests a possible correlation between the mass of the second-born neutron star, $M_{\rm NS,2}$ and the orbital eccentricity $e$ in double neutron star systems \citep{Tauris+2017ApJ...846..170T}. 

To examine this correlation, we compiled $M_{\rm NS,2}$ and $e$ values\textsuperscript{\ref{NSmass}} from known Galactic DNS systems: PSRs~J0453+1559 \citep{Martinez+2015ApJ...812..143M}, J0509+3801 \citep{McEwen+2024ApJ...962..167M}, J0737--3039A/B \citep{Kramer+2021PhRvX..11d1050K}, J1518+4904 \citep{Tan+2024ApJ...966...26T}, B1534+12 \citep{Fonseca+2014ApJ...787...82F}, J1756--2251 \citep{Ferdman+2014MNRAS.443.2183F}, J1757--1854 \citep{Cameron+2023MNRAS.523.5064C}, J1829+2456 \citep{Haniewicz+2021MNRAS.500.4620H}, J1913+1102 \citep{Ferdman+2020Natur.583..211F}, B1913+16 \citep{Weisberg+2016ApJ...829...55W}, and J1946+2052 \citep{Meng+2025arXiv251012506M}.
% and B2127+11C \citep{Jacoby+2006ApJ...644L.113J}.
As shown in Fig.~\ref{MNS-ECC}, a trend is visible between $M_{\rm NS,2}$ and orbital eccentricity. The position of PSR~J0641+0448 which is marked by a red star, shows a low companion mass and moderate eccentricity,  consistent with the proposed correlation.

%% Please use the acknowledgment and contribution environments. This will 
%% be anonomyized when the "anonymous" style option is used. 
\begin{acknowledgments}
This work made use of the data from FAST (https://cstr.cn/31116.02.FAST). FAST is a Chinese national mega-science facility, built and operated by the National Astronomical Observatories, Chinese Academy of Sciences. We all appreciate the excellent performance of the FAST and the operations team. 
The authors were supported by the  National Key R\&D Program of China No. 2025YFA1614000, National Natural Science Foundation of China (NSFC), grant numbers 12588202 and 12041303, the Chinese Academy of Sciences via project JZHKYPT-2021-06, 
and the National SKA Program of China grant 2020SKA0120100 and 2022SKA0120103.
\end{acknowledgments}

\begin{contribution}
%%This section gives authors the space to recognize author contributions. The text inside this environment is NOT counted towards the total word quanta. At a minimum, manuscripts are expected to include this text:

% All authors contributed equally to the Terra Mater collaboration.

%% But authors are expected to provide more specific details, e.g. 
%%
%%SC was responsible for writing and submitting the manuscript.
%%WWM came up with the initial research concept and edited the manuscript.
%%OTS obtained the funding and edited the manuscript.
%%EBF provided the formal analysis and validation. He also edited the manuscript.
%%GEH Supervised the undergraduates, wrote the software and administers the project github and Zenodo repositories.
%%
%% Authors can use the Contributor Role Taxonomy (CRediT) at
%% for ideas on how write a good statement tailored to their needs.

\end{contribution}

\facilities{FAST}

\software{Presto \citep{Ransom+2001PhDT.......123R},  Dspsr \citep{Straten+2011PASA...28....1V},
Psrchive \citep{Hotan+2004PASA...21..302H}
          }


\appendix

%% For this sample we use BibTeX plus aasjournalv7.bst to generate the

% \bibliography{ref}{}
\bibliographystyle{aasjournalv7}



\end{document}